# Broken translational and rotational symmetries in LiMn$_{1.5}$Ni$_{0.5}$O$_4$ spinel

Birender Singh[1], Sunil Kumar[2] and Pradeep Kumar[1*]

[1]School of Basic Sciences, Indian Institute of Technology Mandi, Mandi-175005, India

[2]Discipline of Metallurgy Engineering and Materials Science, Indian Institute of Technology Indore, Simrol, Indore-453552, India

**Abstract:** In condensed matter physics broken symmetries and emergence of quasi-particles are intimately linked to each other. Whenever a symmetry is broken, it leaves its fingerprints, and that may be observed indirectly via its influence on the other quasi-particles. Here, we report the strong signature of broken rotational symmetry induced due to long range-ordering of spins in Mn - sublattice of LiMn$_{1.5}$Ni$_{0.5}$O$_4$ below $T_c$ ~ 113 K reflected with the marked changes in the lattice vibrations using Raman scattering. In particular, the majority of the observed first-order phonon modes show a sharp shift in frequency in the vicinity of long range magnetic-ordering temperature. Phonons exist in a crystalline system because of broken translational symmetry, therefore any renormalization in the phonon-spectrum could be a good gauge for broken translational symmetry. Anomalous evolution of the few modes associated with stretching of Mn/NiO$_6$ octahedra in the intermediate temperature range (~ 60-260 K) marked the broken translational symmetry attributed to the charge ordering. Interestingly same modes also show strong coupling with magnetic degrees of freedom, suggesting that charge-ordering and magnetic transition may be linked to each other.

*email id: pkumar@iitmandi.ac.in

# INTRODUCTION

Since last few decades, numerous research efforts have been made in the field of advanced rechargeable batteries because of their potential applications in energy storage for electronics and portable devices (i.e. cell phones, laptop, electric vehicles etc.). Lithium-ion batteries appeared as the promising candidate due to its high energy density and high electrochemical performance [1-6]. $LiMn_2O_4$ spinel material has attracted significant research interest as a cathode material for fabrication of high capacity Li-ion batteries due to its stable three-dimensional structure for lithium intercalation, inexpensive and environmentally suitable as compared to earlier commercialized layered material, i.e. $LiCoO_2$ and $LiNiO_2$ [7-9]. Although, $LiMn_2O_4$ a manganese spinel emerges as promising cathode material for Li-ion batteries but found to be compromise in electrochemical performance. In the stoichiometric $LiMn_2O_4$, manganese exhibits both +3 and +4 (i.e. $Mn^{3+}$ ($t_{2g}^3 - e_g^1$) and $Mn^{4+}$ ($t_{2g}^3 - e_g^0$)) oxidation state, where the presence of $Mn^{3+}$ ($t_{2g}^3 - e_g^1$) in local atomic structural configuration produces octahedral distortion due to Jahn-Teller (JT) effect, however, $Mn^{4+}$ ($t_{2g}^3 - e_g^0$) is free from JT effect [10-11]. Due to the presence of JT-distortion, it undergoes a charge-ordered reversible structural phase transition from cubic to orthorhombic phase around room temperature associated with orbital-ordering [12-15] and results in reduced charge-discharge cycling performance. However, the partial substitution of different transition-metal ions (i.e. Fe, Co, Cr, Al and Ni) on Mn-site produces remarkable features with suppression in JT-distortion [16-21] and enhanced local structural stability. The fractional substitution of $Ni^{2+}$ ($t_{2g}^6 - e_g^2$) in Mn-site of $LiMn_2O_4$ spinel results in increased rate performance due to the enormous change in $Mn^{3+}/Mn^{4+}$ ratio [22-24]. In addition, $LiMn_{1.5}Ni_{0.5}O_4$ has a higher operating potential ~ 4.7 V *versus* $Li^+/Li$ as compared to the parent $LiMn_2O_4$. In spinel $LiMn_{1.5}Ni_{0.5}O_4$, manganese exhibit stable average +4 oxidation state, which stabilizes the three-dimensional structure and results in improved electrochemical charge/discharge cycling performance and thermal stability. $LiMn_{1.5}Ni_{0.5}O_4$ spinel is known to crystallize in two different phases (i.e. ordered $P4_332$-phase and disordered $Fd\bar{3}m$-phase) at room temperature depending on the synthesis conditions. In ordered phase, $Ni^{2+}$ and $Mn^{4+}$ ions occupy different octahedral sites (4a and 12d, respectively); whereas, in disordered phase they occupy the same 16d-octahedral site. It is crucial to investigate the microscopic peculiarities to understand the underlying physics in this compound. As phonons are responsible for controlling many physical properties of a system, such as thermal/electronic transport and stability of a system, therefore it becomes very important to probe the dynamics of phonons and its coupling with other quasiparticles in this system.

Here, we present a comprehensive structural, magnetic and inelastic light (Raman) scattering studies of $LiMn_{1.5}Ni_{0.5}O_4$ along with first principle phonon calculations of $LiMn_2O_4$ to carefully investigate the local structural changes as well as coupling of phonons with spin, charge and/or orbital excitations. Interestingly, we observed a significant mode hardening at ~ 60 K in prominent vibrational modes S1 ( ~ 167 $cm^{-1}$), S7 ( ~ 505 $cm^{-1}$), S9 ( ~ 600 $cm^{-1}$) and S11 ( ~ 644 $cm^{-1}$) attributed to the presence of long range spin ordering below this temperature. However, the phonon mode S10 ( ~ 616 $cm^{-1}$) associated with Mn-O vibration exhibit anomalous phonon softening and linewidth broadening below 60 K indicating its strong interaction with spin degrees of freedom within the $Mn^{4+}$ sublattice. Furthermore, we noticed significant evolution of S10 mode ( ~ 616 $cm^{-1}$) below ~ 260 K strongly indicate charge-ordering on cooling, interestingly only in the intermediate temperature range of ~ 60-260 K. The anomalous temperature dependent behavior of phonon mode frequencies and line-width reflects the presence of strong spin-lattice interaction and charge-ordering in $LiMn_{1.5}Ni_{0.5}O_4$ spinel at low temperature.

**Experimental and Computational details**

LiMn$_{1.5}$Ni$_{0.5}$O$_4$ sample was synthesized via a sol-gel process. Stoichiometric amounts of LiNO$_3$, Ni(NO$_3$)$_2$.6H$_2$O, and Mn(NO$_3$)$_2$ (all purity > 99.9%) precursors were dissolved in de-ionized water separately and were mixed while stirring vigorously for uniform mixing. A certain amount of solution of citric acid and ethylene glycol (in 1:1 molar ratio) was added to the solution containing the metal ions. This solution was heated at 80 ºC for 5 h, and subsequently dried at 120 ºC. The resultant powder was hand ground using a mortar & pestle and calcined at 400 ºC for 6 h, and then at 750 °C for 12 h. A Bruker D2 Phaser X-ray diffractometer equipped with Cu K$_\alpha$ as the x-ray source ($\lambda = 1.54$ Å) was used for the crystal structure analysis. The x-ray diffraction data were collected in the 2θ range of 10° - 80°. Raman spectroscopic measurements were carried out in a broad temperature range (5 - 400 K) using Raman spectrometer similar to as described in Ref. [25] and [26]. The temperature and dc magnetic field dependent magnetization measurements were carried out with a Quantum Design (USA) Magnetic Property Measurements System (MPMS).

The first principle calculations were done for parent compound via using ultrasoft pseudopotential plane-wave approach within the framework of the density functional theory (DFT) through QUANTUM ESPRESSO package [27]. The dynamical matrix and phonons at gamma point (q= 0, 0, 0) were calculated using Perdew-Burke-Ernzerhof generalized gradient approximation [28] as exchange-correlation functional along with linear response approach within density functional perturbation theory [29]. In these calculations, we have used the plane wave energy cut-off of 60 Ry and the charge density cut-off of 280 Ry to expand the wave function. The k-point grid 8 x 8 x 8 is employed for numerical integration of the Brillouin zone (BZ) using Monkhorst-Pack [30].

# RESULTS AND DISCUSSION

## 1. Crystal Structure

Figure 1 shows the room temperature powder x-ray diffraction (XRD) pattern of the LiMn$_{1.5}$Ni$_{0.5}$O$_4$ sample synthesized at 750 °C. Rietveld refinement analysis of the XRD data confirm that no impurity or secondary phase is present in the sample. An excellent match between the observed diffraction pattern (open symbols) and calculated diffraction pattern (thick red line) further confirm that LiMn$_{1.5}$Ni$_{0.5}$O$_4$ sample prepared in present investigation crystallizes in a cubic structure (space group $Fd\bar{3}m$). An illustration of the LiMn$_{1.5}$Ni$_{0.5}$O$_4$ crystal structure is given as the inset in Figure 1. The cell parameter *a* and cell volume *V* was found to be 8.17365(15) Å and 546.069(30) Å$^3$, respectively which is in agreement with the earlier reports [31].

## 2. Magnetic Study

Figure 2 (a) shows temperature variation zero field cooled (ZFC) and field cooled (FC) magnetic susceptibility ($\chi$) of LiMn$_{1.5}$Ni$_{0.5}$O$_4$ spinel in the temperature range of 2-300 K at applied dc magnetic field of 100, 500 and 1000 Oe. The $\chi$ measurement shows abrupt increase below ~ 115 K on cooling and significant bifurcation between ZFC and FC curves is observed at low applied dc magnetic field ( 100 Oe ) with a broad maximum cusp in ZFC curve around 50 K. The observed feature of $\chi$ indicates the ferrimagnetic interaction within Mn and Ni sub-lattices with the presence of spin-glass magnetic nature in this material [32]. Furthermore, the bifurcation between ZFC and FC curves substantially decreases with an increase in the strength of the applied dc magnetic field at 500 Oe and 1000 Oe (see Figure 2 (a))). The inverse magnetic susceptibility ($\chi^{-1}$) data (not shown) at 100 Oe was fitted with Curie-Weiss (CW) law in the temperature range of 110 to 300 K. The estimated value of effective magnetic moment ($\mu_{eff}$) and Curie-Weiss temperature ($\theta_{cw}$) obtained from fitting is ~ 3.5 ($\mu_{eff}/f.u.$) and ~ 113 K, respectively. The positive value of $\theta_{CW}$ suggests the increased ferromagnetic contribution in this Ni-doped spinel compound.

Figure 2 (b) shows isothermal magnetization M (*H*) curves of LiMn$_{1.5}$Ni$_{0.5}$O$_4$ at different fixed temperature between 2 to 300 K and inset display nature of magnetization M (*H*) at the low field region at 2 and 60 K. Magnetization plot shows linear magnetic field dependence down to 150 K and below this temperature clear non-linear behavior in magnetization is observed. Furthermore, magnetic hysteresis associated with non-linear plot is observed to build up gradually on decreasing temperature below 150 K and show sufficient increase at low temperature (see inset Figure 2 (b)). The observed low temperature features (i.e., non-linear behaviour of magnetization and the presence of magnetic hysteresis below ~ 150 K) are attributed to increased ferromagnetic interaction within the sub-lattice [33-34]. At low temperature, magnetization shows an abrupt increase to its maximum value and get saturated at very low value of applied dc magnetic field ( < 0.5 kOe). In LiMn$_{1.5}$Ni$_{0.5}$O$_4$ spinel, as manganese generally exhibit +4 oxidation state (Mn$^{4+}$; $t_{2g}^3 - e_g^0$ orbital configuration) and nickel has +2 oxidation state (Ni$^{2+}$; $t_{2g}^6 - e_g^2$ orbital configuration), therefore, provide two different types of spin interaction in octahedral sub-lattice structure. First, a ferromagnetic super-exchange interaction having approximately 90º angle as: Mn$^{4+}$-O$^{2-}$-Mn$^{4+}$, Ni$^{2+}$-O-Ni$^{2+}$, and second corresponds to Mn$^{4+}$ - O - Ni$^{2+}$ antiferromagnetic ordering [35-37]. The strongly competing spin-interaction between these two magnetic sub-lattices gives rise to collinear ferrimagnetic spin alignment below ~ 113 K, where spin associated with Mn$^{4+}$ aligns along the magnetic field and Ni$^{2+}$ spins align in the opposite direction.

## 3. Inelastic Light ( Raman ) Scattering Study
### 3.1 Group theory analysis and mode assignments

LiMn$_{1.5}$Ni$_{0.5}$O$_4$ crystallizes in ordered (space group $P4_332$) as well as disordered (space group $Fd\bar{3}m, O_h^7$) phase [31, 38]. In accordance with the factor group analysis irreducible representation of order and disordered phase of LiMn$_{1.5}$Ni$_{0.5}$O$_4$ spinel consists of 168 and 42 phonon modes, respectively [39]. Total number of Raman and infrared (IR) active modes in order and disordered phase within irreducible representations and corresponding mode decomposition is summarized in Table I. Figure 3 illustrate the Raman spectrum in the spectral range of 100-800 cm$^{-1}$ obtained at 5 K, highlighting total 12 phonon modes named as S1 to S12. The observed Raman spectrum shows the presence of broad low intensity modes around ~ 400 cm$^{-1}$ suggesting the disordered nature of LiMn$_{1.5}$Ni$_{0.5}$O$_4$ ($Fd\bar{3}m$), as these modes were observed to be relatively intense in ordered-$P4_332$ phase [40-41]. The spectra are fitted with a sum of Lorentzian to determine the self-energy parameter such as phonon mode frequency and line-width. All the observed modes S1 to S12 at 5 K are tabulated in Table-II. Furthermore, to understand the microscopic behavior of vibrational modes associated with LiMn$_{1.5}$Ni$_{0.5}$O$_4$, we have performed zone-centred phonon calculations for the undoped compound. Our density functional perturbation theory calculated phonon mode frequencies are in good agreement with the experimentally observed values [42] (see Table-II). We have compared the calculated mode frequencies with the experimentally observed frequencies at 5 K as: $|\bar{\omega}_r|_\% = \frac{100}{N}\sum_i \left|\frac{\omega_i^{cal} - \omega_i^{exp}}{\omega_i^{exp}}\right|$, where $i = 1, 2 ----, N$ is the number of Raman active modes (5 here) and $|\bar{\omega}_r|_\%$ is the average of the absolute relative differences. The obtained value of $|\bar{\omega}_r|_\%$ in the mode frequencies is observed to be very nominal i.e. ~ 1%. Figure 4 illustrates the calculated eigen vector for the phonon modes S4, S5, S7, S10 and S11, indicating the direction and magnitude of atomic vibration along with the atoms participating in a particular mode of vibration. Based on our first-principle calculations, we have assigned the observed phonon modes as: S2-S3 ( ~ 332-345 cm$^{-1}$; $F_{2g}^{(3)}$), S4-S5 ( ~ 400-416 cm$^{-1}$; $E_g$ ) modes arise due to the vibration of O-atoms in the tetrahedral/octahedral coordination; the most intense phonon mode S7 ( ~ 505 cm$^{-1}$; $F_{2g}^{(2)}$ ) and S9-S10 ( ~ 600 - 616 cm$^{-1}$ $F_{2g}^{(1)}$ ) are associated with motion of Li and O atoms, and S11 ( ~ 644 cm$^{-1}$; $A_{1g}$ ) originates from the symmetric stretching vibration of Mn-O bond in the MnO$_6$ octahedra. In addition, the appearance of low frequency mode S1 ~ 167 cm$^{-1}$ in the spectrum may be speculated due to the scattering of the acoustic mode from other parts of Brillouin zone, as the phonon dispersion of LiMn$_2$O$_4$ spinel shows the value close to this frequency at X-point in the Brillouin zone [43]. Also, the observed weak shoulder mode S12 may be attributed to the second order modes or second order IR modes become Raman active [44-45].

### 3.2 Temperature Dependence of Phonon Modes

Figure 5 shows the temperature evolution of the self-energy parameters of the phonon modes (i.e. S1, S4 and S5) in the spectral range of 100 - 450 cm$^{-1}$. We note that frequency of the mode S1, S4 and S5 start deviating from normal temperature dependence i.e. increase in mode frequency with decreasing temperature below ~ 110 K; and interestingly, S1 shows a small jump in frequency at ~ 60 K and remain constant below this temperature. However, damping constant shows normal temperature evolution i.e. decrease in linewidth with decreasing temperature.

Figure 6 illustrates the temperature dependence of mode frequency and line-width of the prominent phonon modes (i.e. S7, S9-S11) in the spectral range of 450-800 cm$^{-1}$. Following observations can be made: (i) frequency of mode S9 and the most prominent modes i.e. S7 and S11 shows sudden jump in the frequency at ~ 60 K and below this temperature frequency remains almost constant down to 4 K. On the other hand frequency of mode S10 shows softening below ~ 60 K; (ii) line-width of S7 and S11 modes start deviating from their normal temperature behavior below ~ 110

K. However, line-width of the mode S9 and S10 shows anomalous temperature evolution; first their linewidth decreases sharply till ~ 230 K starting from 400 K and then line-width of S9 changes slope. Linewidth of S10 is nearly constant below 230 K down to ~ 100 K, and on further cooling continuous line broadening (~ 25 %) is observed down to lowest recorded temperature (see Figure 6). Figures 7 ( a and b ) shows the temperature evolution of S9 and S10 modes. Above room temperature S10 mode is weak and appeared as a shoulder of S9 mode. Interestingly, with lowering temperature below ~ 260 K, it shows distinct evolution down to ~ 60 K, and on further cooling, it again starts merging and becomes a shoulder-like feature of S9 mode. The distinct evolution of S10 mode around ~ 610 $cm^{-1}$ in the intermediate temperature range of 60-260 K is attributed to the improved Mn/Ni charge-ordering [46]. Interestingly, the intensity of S9 and S10 mode also shows anomalous temperature dependence (see Figure 7 (c and d)). Intensity of the S9 mode increases with decrease in temperature down to ~ 230 K and below this temperature it is observed to remain nearly constant down to 5 K. On the other hand, intensity of S10 mode remain almost constant down to ~ 230 K and increases on further decrease in temperature upto ~ 100 K, below this temperature it is again nearly constant down to 5 K. The anomalous temperature dependent behavior of intensity associated S9 and S10 mode again reflect the appearance of charge-ordering (broken translational symmetry) in $LiMn_{1.5}Ni_{0.5}O_4$ spinel. The anomaly observed in the line-width of S9 and S10 mode at 230 K may have its origin due to the interaction of phonons with charge degrees of freedom below ~ 230 K down to 60 K (see Figure 6). This local charge-ordering in B-site of spinel is associated with $Mn/NiO_6$ octahedral arrangement, as modes around 600 $cm^{-1}$ arise mainly due to Mn-O stretching vibration. The anomalous temperature variation of intensity may also be interpreted as a direct consequence of the presence of distorted and undistorted $MnO_6$ octahedra in the structural unit cell, indicating the presence of a small amount of $Mn^{3+}$ ions in this spinel as manganese in +3 oxidation-state is responsible for octahedral distortion known as Jahn-Teller distortion effect. The partial charge-ordering of $Mn^{3+}$ and $Mn^{4+}$ ions is also observed in parent $LiMn_2O_4$ manganese-spinel in orthorhombic phase below 280 K [47-50]. Similar charged-ordering of $Fe^{3+}$ and $Fe^{4+}$ ions observed in magnetite $Fe_3O_4$ is viewed as a Verwey transition [51]. Strong interaction between spin and lattice degrees of freedom, and associated charged and/or orbital ordering has also been observed in similar spinel compounds in low temperature range leading to magnetic and/or structural phase transition [52-53]. Nevertheless, our study shows a significant effect of spin and/or charge degrees of freedom on lattice dynamics of $LiMn_{1.5}Ni_{0.5}O_4$ compound at low temperature reflecting the complex interplay between various degrees of freedom. Furthermore, the anomalous temperature dependence of mode frequencies and line-width at low temperature (< 60 K) reflect strong interaction of phonon with magnetic excitations.

At low temperature, the intrinsic lattice dynamics has a small contribution to any change in the frequency and line-width (or inverse of life-time) of phonon mode. Within the harmonic approximation of phonon dynamics, zone centre phonons (i.e. optical phonons, $\vec{q} = 0$) frequency do not show variation with temperature and have infinite life-time; however, the real system is perturbed with anharmonicity. Anharmonic phonon-phonon scattering consequently gives rise to decay of high energy optical phonon ($\vec{q} = 0$) into two lower energy acoustic phonons of opposite momentum and equivalent energy. This decay channel of optical phonon ($\vec{q} = 0$) due to anharmonic coupling is primarily responsible for the temperature variation of phonon frequency ($\omega$) and non-zero line-width (i.e. finite life-time) of phonon modes in Raman spectra. In order to obtain insight into the temperature dependence of phonon modes, we have fitted line-width ($\Gamma$) as [54]: $\Gamma(T) = \Gamma_0 [1 + (2\lambda_{ph-ph})/(e^x - 1)]$, where, $x = \hbar\omega_0 / 2K_B T$ and $\omega_0$, $\Gamma_0$ are the mode frequency and line-width, respectively at absolute zero temperature, and $\lambda_{ph-ph}$ corresponds to phonon-phonon coupling parameter. Temperature variation of phonon mode frequency ($\omega$) depends as follows [55]:

$$\omega(T) = \omega_0 + (\Delta\omega)_{latt}(T) + (\Delta\omega)_{anh}(T) + (\Delta\omega)_{el-ph}(T) + (\Delta\omega)_{sp-ph}(T) \quad (1)$$

In this equation, second term $(\Delta\omega)_{latt}(T)$ corresponds to change in phonon frequency due to lattice thermal expansion, termed as the quasi-harmonic effect, given as

$$(\Delta\omega)_{latt}(T) = \omega_0 \left( \exp\left(-3\gamma_0 \int_0^T \alpha(T)\, dT\right) - 1 \right) \quad (2)$$

where $\gamma_0$ is the Gruneisen parameter and $\alpha(T)$ is linear thermal expansion coefficient. For simplicity, we have written the product of $\gamma_0$ and $\alpha(T)$ as a series of temperature, given as

$$3\gamma_0 \alpha(T) = a_0 + a_1 T + a_2 T^2 + a_3 T^3 + .... \quad (3)$$

where $a_0, a_1, a_2$ and $a_3$ are constant. Third term in equation (1) arises from anharmonic phonon-phonon interaction (i.e. decay of phonon), given as: $(\Delta\omega)_{anh}(T) = -A[1 + 4\lambda_{ph-ph}/e^x - 1]$ where, $A = \Gamma_0^2/2\omega_0$, and last two terms of eqn. (1) are attributed to change in phonon frequency due to electron-phonon coupling and spin-phonon coupling, respectively, and these can be neglected as this spinel compound is insulating in nature and shows no magnetic correlation above ~ 113 K. As a result, we have fitted all the phonon modes in the temperature range 120 - 400 K and the obtained fitted parameters are summarized in Table-II. The value of phonon-phonon coupling ($\lambda_{ph-ph}$) is found to be maximum for S10 ( $F_{2g}^{(1)}$ ) ( $\lambda_{ph-ph}$ ~ 1 ) and followed by S11 ( $A_{1g}$ ) (~ 0.6 ) and S7 ( $F_{2g}^{(2)}$ ) (~ 0.5 ) modes indicate that strong phonon-phonon interaction is primarily prominent in modes arising due to stretching of Mn-O and Ni-O bonds.

The observed magnetic ordering is reflected as the deviation of mode frequency and linewidth from estimated temperature variation via phonon-phonon interaction model ( see solid red curve in Figure 5 and 6 ) below the magnetic transition due to the interaction between lattice and magnetic degrees of freedom through a strong spin-phonon coupling. We note that some of the modes show pronounced softening/hardening as compared to other phonon modes below the magnetic ordering temperature. Microscopically, we may understand the different nature of coupling with spin degrees of freedom through the exchange Hamiltonian for nearest-neighbours spin-spin correlation given as [56-58]:

$$H = \frac{1}{2}\sum_{ij} J_{ij} <\vec{S_i}.\vec{S_j}> \tag{4}$$

where, $J_{ij}$ is a super-exchange integral, $<\vec{S_i}.\vec{S_j}>$ is the spin-spin correlation function. The amplitude of spin-spin correlation function $<\vec{S_i}.\vec{S_j}>$ is responsible for the phonon modulation of the super-exchange integral and is reflected as strong spin-phonon coupling at low temperature. Phonon renormalization is directly related to super-exchange integral and spin-spin correlation function given as,

$$\Delta\omega = \frac{1}{2m\omega}\sum_{ij} \frac{\partial^2 J_{ij}}{\partial u^2} <\vec{S_i}.\vec{S_j}> \tag{5}$$

where, m, $\omega$ and u represent mass, unrenormalized mode frequency and displacement of atom from their mean position, respectively. The exchange integral $J_{ij}$ can be positive or negative depending on antiferromagnetic and ferromagnetic interactions, respectively. Also, it depends on the displacement of atoms/ions participating in a particular phonon mode. As most of the phonon modes show significant hardening below the magnetic ordering temperature indicate that standard anharmonic behaviour is not followed in this regime. The observed phonon modulation can be attributed to increased amplitude of spin-spin correlation function $<\vec{S_i}.\vec{S_j}>$ and net increased in spin-phonon coupling via modulation of the exchange parameter $J_{ij}$. The amount of deviation of phonon mode frequency (i.e. $\Delta\omega$) from normal anharmonic interaction model depends on the strength of coupling of a particular phonon mode with underlying magnetic degrees of freedom. As different phonon modes arise due to the vibration of different atoms with varying amplitude and direction of vibration (see Figure 4; where the length of arrow shows amplitude of vibration associated with particular atom), therefore the strength of spin-phonon coupling may vary in accordance to particular mode of vibration; and varying renormalization may be attributed to these factors.

As it is well understood that the structural stability is very crucial for better electrochemical performance during the charge/discharge cycle (i.e. during lithium intercalation / deintercalation). Therefore, it is important to have better insight into the structural stability of this spinel compound. We have studied the temperature variation of linear thermal expansion coefficient $(\alpha(T))$ extracted from our Raman measurements using eqn. 3, where the value of constants $a_0, a_1, a_2$ and $a_3$ are obtained from fitting as discussed above in the temperature range of 120-400 K. However, below 120 K, it is only estimated value of $\alpha(T)$ (see Figure 8) based on the constant extracted in the temperature range of 120-400 K; and value of the Gruneisen parameter is taken as ~ 2. Figure 8 (a, b and c) shows the variation of $\alpha(T)$ for predominant phonon modes (i.e. S4, S5, S7, S9, S10 and S11). The following observations can be made: (i) the temperature variation of $\alpha(T)$ associated with low frequency modes S4 and S5 decreases on cooling down to lowest recorded temperature; however, the observed decrease in $\alpha(T)$ is more prominent below ~ 110 K (see Figure 8 (a)); (ii) $\alpha(T)$ for S9 decreases with decreasing temperature down to ~ 5 K; on the other hand for S10 mode shows slight upward trend below ~ 110 K (see Figure 8 (b)) and (iii) for the most intense modes S7 and S11, the value of $\alpha(T)$ decrease on cooling down to ~ 110 K and on further cooling it is found to be nearly constant for S7 mode and sharp decrease is observed for S11 mode (see Figure 8 (c)). The observed temperature variation of $\alpha(T)$ clearly shows that LiMn$_{1.5}$Ni$_{0.5}$O$_4$ spinel is structurally stable, indicating suppression JT distortion effect (i.e. significantly reduced Mn$^{3+}$ oxidation state) in this compound. As in the pristine LiMn$_2$O$_4$ spinel, the octahedral distortion due to the presence of JT effect causes first order structural phase transition and is reflected as a relatively increased value of $\alpha(T)$ around the transition temperature [59].

## SUMMARY AND CONCLUSIONS

In conclusion, we have carried out magnetic, inelastic light (Raman) scattering study of LiMn$_{1.5}$Ni$_{0.5}$O$_4$, and density functional theory based zone centred phonon calculations. Our magnetization study suggests the competing interaction between two different magnetic sub-lattices associated with Mn$^{4+}$/Ni$^{2+}$O$_6$ octahedral, result in ferrimagnetic ordering below ~ 113 K with an effective magnetic moment $\mu_{eff}$ ~ 3.5 ($\mu_{eff}/f.u.$). The inelastic light (Raman) scattering study demonstrate the symmetry breaking, anticipated to the interaction between different degrees of freedom such as lattice, spin and charge, reflected via the anomalous softening/hardening of different phonon modes and observed anomaly in the temperature variation of line-widths and intensity of modes associated with vibration of Mn/NiO$_6$ octahedra. The temperature variation of $\alpha(T)$ suggests that this spinel compound is structurally more stable. Furthermore, the density functional theory calculated zone cantered phonon frequencies for parent compound are found to be in very good agreement experimental values. The present results provide a critical understanding about the microscopic phenomenon of this spinel compound through lattice dynamics study and further call for exploring the charge-ordering phase within the intermediate temperature regime.

## Acknowledgements

PK and SK acknowledge the Department of Science and Technology, India, for the grant under INSPIRE faculty fellowship. PK acknowledges Advanced Material Research Centre, IIT Mandi, for the experimental facilities.

**Table-I:** Wyckoff positions and irreducible representations of phonon modes for $LiMn_{1.5}Ni_{0.5}O_4$ in ordered (space group, $P4_332$) and disordered (space group, $Fd\bar{3}m$) phase.

| Ordered $P4_332$-Phase | | | |
|---|---|---|---|
| Atom | site | Wyckoff position | Mode decomposition |
| Li | $C_3$ | 8c | $A_1 + A_2 + 2E + 3F_2 + 3F_1$ |
| Mn | $C_2$ | 12d | $A_1 + 2A_2 + 3E + 4F_2 + 5F_1$ |
| Ni | $D_3$ | 4a | $A_2 + E + F_2 + 2F_1$ |
| O1 | $C_3$ | 8c | $A_1 + A_2 + 2E + 3F_2 + 3F_1$ |
| O2 | $C_1$ | 24e | $3A_1 + 3A_2 + 6E + 9F_2 + 9F_1$ |
| Raman / Infrared Active modes | | $\Gamma_{Raman} = 6A_1 + 14E + 20F_2$; $\Gamma_{IR} = 21F_1$; $\Gamma_{Acoustic} = F_1$ | |
| Disordered $Fd\bar{3}m$-Phase | | | |
| Atom | Site | Wyckoff position | Mode decomposition |
| Li | $T_d$ | 8a | $F_{2g} + F_{1u}$ |
| Mn/Ni | $D_{3d}$ | 16d | $A_{2u} + E_u + F_{2u} + 2F_{1u}$ |
| O | $C_{3v}$ | 32e | $A_{1g} + A_{2u} + E_u + E_g + F_{2u} + 2F_{2g} + 2F_{1u} + F_{1g}$ |
| Raman / Infrared Active modes | | $\Gamma_{Raman} = A_{1g} + E_g + 3F_{2g}$; $\Gamma_{IR} = 4F_{1u}$; $\Gamma_{Acoustic} = F_{1u}$; $\Gamma_{silent} = F_{2u}$ | |

**Table-II:** Experimentally observed phonon frequencies of LiMn$_{1.5}$Ni$_{0.5}$O$_4$ at 5 K, fitting parameters, fitted using equations as described in the text, and calculated and experimental phonon frequencies of LiMn$_2$O$_4$. (Units are in cm$^{-1}$).

| Mode | Exp. 5 K | DFT $\omega$ | Exp. | $\omega_o$ | $\Gamma_o$ | $A = \Gamma_o^2/2\omega_o$ | $\lambda_{ph-ph}$ |
|---|---|---|---|---|---|---|---|
| S1 | 167.5 | - | | 166.3 | 11.9 ± 0.15 | 0.429 | 0.023 |
| S2 ($F_{2g}^{(3)}$) | 332.7 | - | | 333.9 | 13.1 ± 0.31 | 0.257 | 0.215 |
| S3 ($F_{2g}^{(3)}$) | 345.3 | 378.3 | 382 [42] | - | - | - | - |
| S4 (E$_g$) | 400.8 | - | | 398.7 | 24.8 ± 0.42 | 0.770 | 0.212 |
| S5 (E$_g$) | 416.1 | 415.3 | 430 [42] | 415.6 | 22.1 ± 0.32 | 0.584 | 0.131 |
| S6 | 488.3 | - | | - | - | - | - |
| S7 ($F_{2g}^{(2)}$) | 504.7 | 494.5 | 492 [42] | 505.4 | 13.4 ± 0.07 | 0.177 | 0.472 |
| S8 | 529.9 | - | | - | - | - | - |
| S9 ($F_{2g}^{(1)}$) | 600.2 | 567.5 | 580 [42] | 600.6 | - | - | 0.199 |
| S10 ($F_{2g}^{(1)}$) | 616.5 | - | | 615.9 | - | - | 0.966 |
| S11 (A$_{1g}$) | 644.2 | 616.5 | 625 [42] | 643.8 | 8.8 ± 0.51 | 0.060 | 0.562 |
| S12 | 661.3 | - | | - | - | - | - |

**FIGURE CAPTION**

**FIGURE 1:** (Color online) Room temperature powder x-ray diffraction pattern with the Rietveld refinement fit for LiMn$_{1.5}$Ni$_{0.5}$O$_4$ sample. Open circle symbols represent the observed pattern, the solid red line represents the calculated pattern, the grey line is the difference between the observed and the calculated patterns, and the green vertical bars at the bottom show the Bragg peak positions for the space group $Fd\bar{3}m$.

**FIGURE 2:** (Color online) (a) Zero-field-cooled and field-cooled magnetic susceptibility ($\chi = (M/H)$) of LiMn$_{1.5}$Ni$_{0.5}$O$_4$ as a function of temperature at dc magnetic field of 100 Oe (red in color), 500 Oe (black in color) and 1000 Oe (blue in color); (b) isothermal magnetization M ($H$) curves of LiMn$_{1.5}$Ni$_{0.5}$O$_4$ at temperature 2 (black), 60 (red), 110 (Green), 150 (blue) and 300 K (cyan). Inset: nature of magnetization M ($H$) at the low field region.

**FIGURE 3:** (Color online) Raman spectra of LiMn$_{1.5}$Ni$_{0.5}$O$_4$ at 5 K. Solid thin lines are the fit of individual modes, and the solid thick line shows the total fit to the experimental data. Inset (a) shows the Raman spectra at 330 K.

**FIGURE 4:** (Color online) Schematic showing the eigen vectors for phonon modes. Green, violet, and red spheres represent Li, Mn, and O atoms, respectively. Red arrows on the atoms represent the direction and magnitude of atomic vibrations.

**FIGURE 5:** (Color online) Temperature dependence of mode frequency and linewidth of S1, S4 and S5 modes. Solid lines are fitted lines curve as described in the text.

**FIGURE 6:** (Color online) Tempe rature dependence of mode frequency and linewidth of S7, S9, S10 and S11 modes. Solid lines are fitted lines curve as described in the text and those for linewidth of modes S9 and S10 are a guide to the eye.

**FIGURE 7:** (Color online) (a and b) Temperature evolution of S9 and S10 modes and (c and d) temperature dependence of the intensity of S9 and S10 modes, solid lines are a guide to the eye.

**FIGURE 8:** (Color online) Temperature dependence of linear thermal expansion coefficient for (a) S4 and S5, (b) S9 and S10, and (c) S7 and S11 modes.

**Figure: 1**

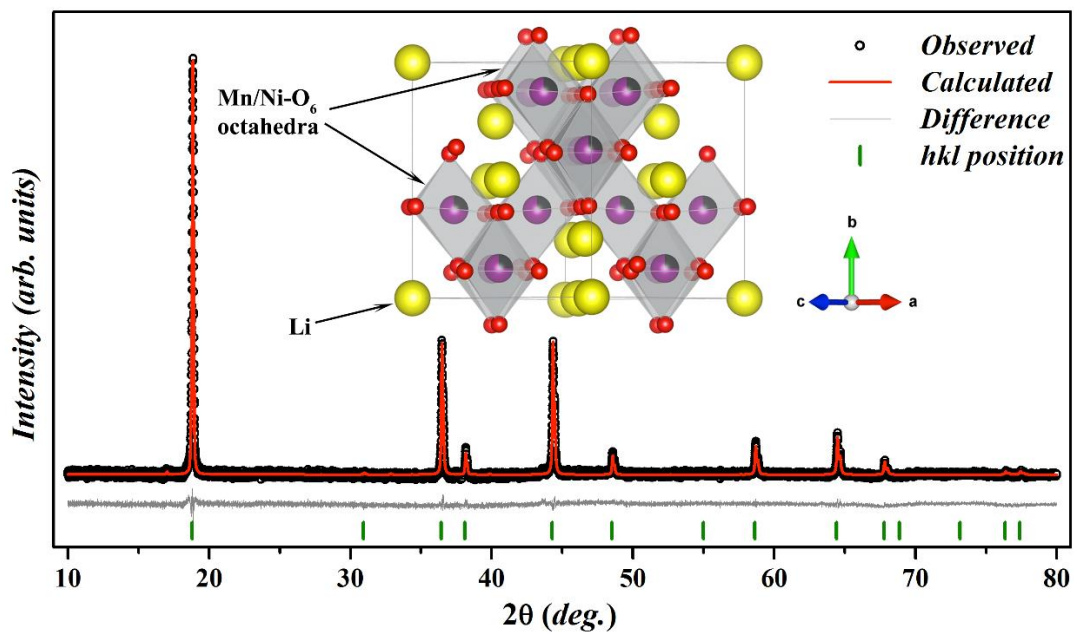

**Figure: 2**

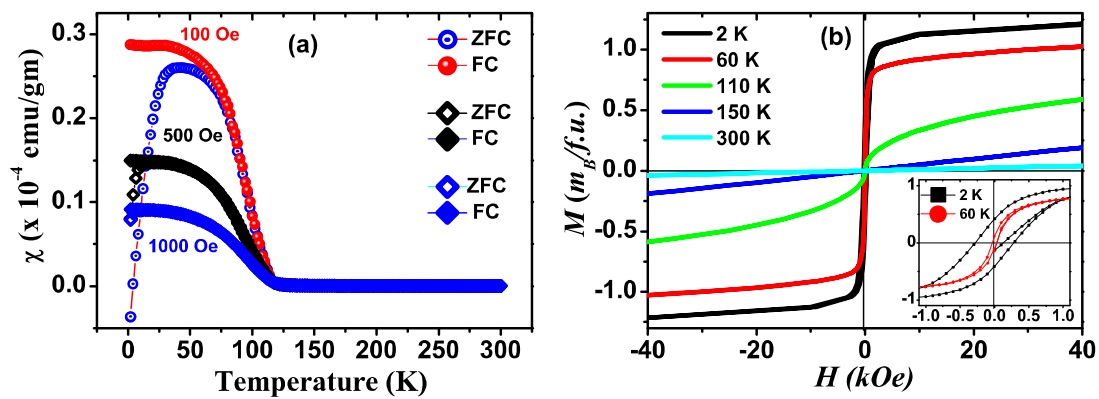

**Figure: 3**

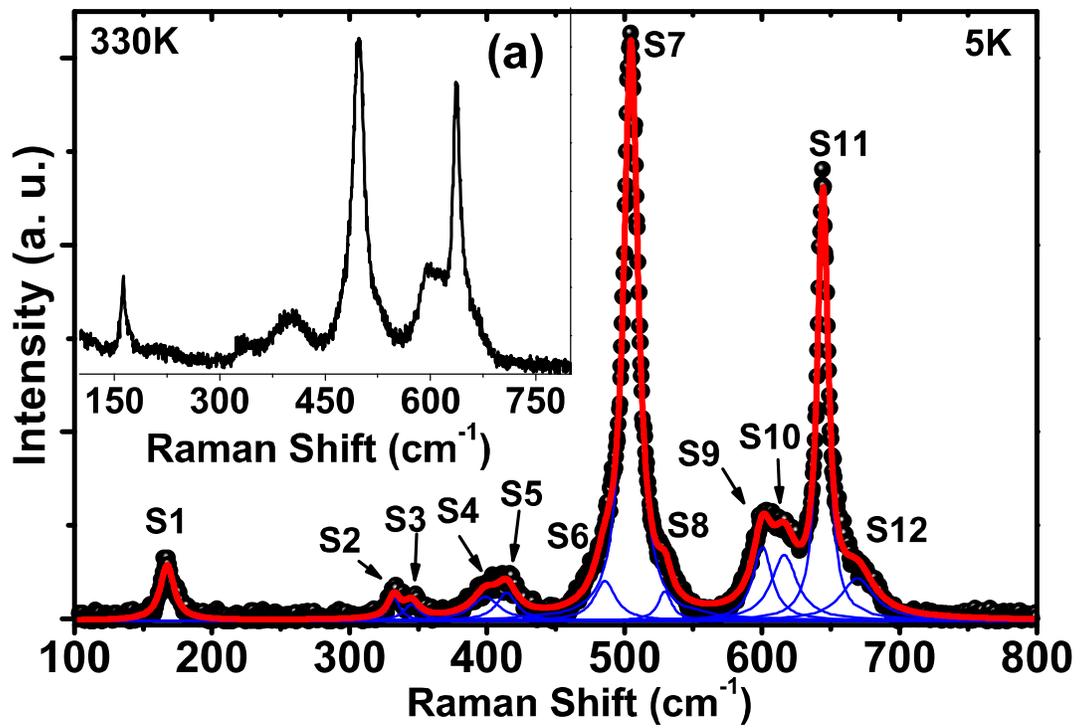

**Figure: 4**

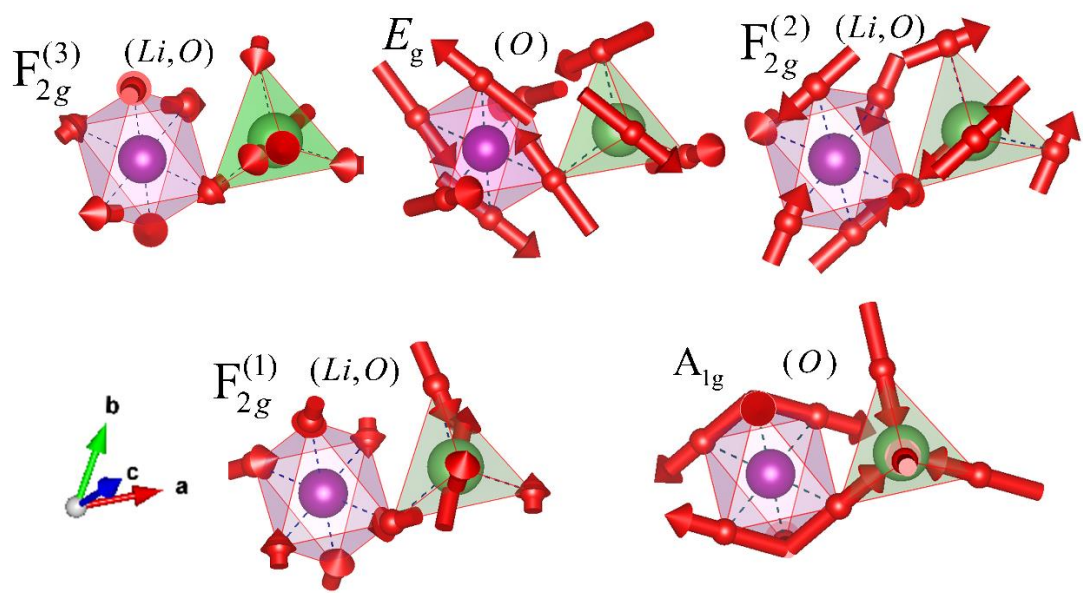

Figure: 5

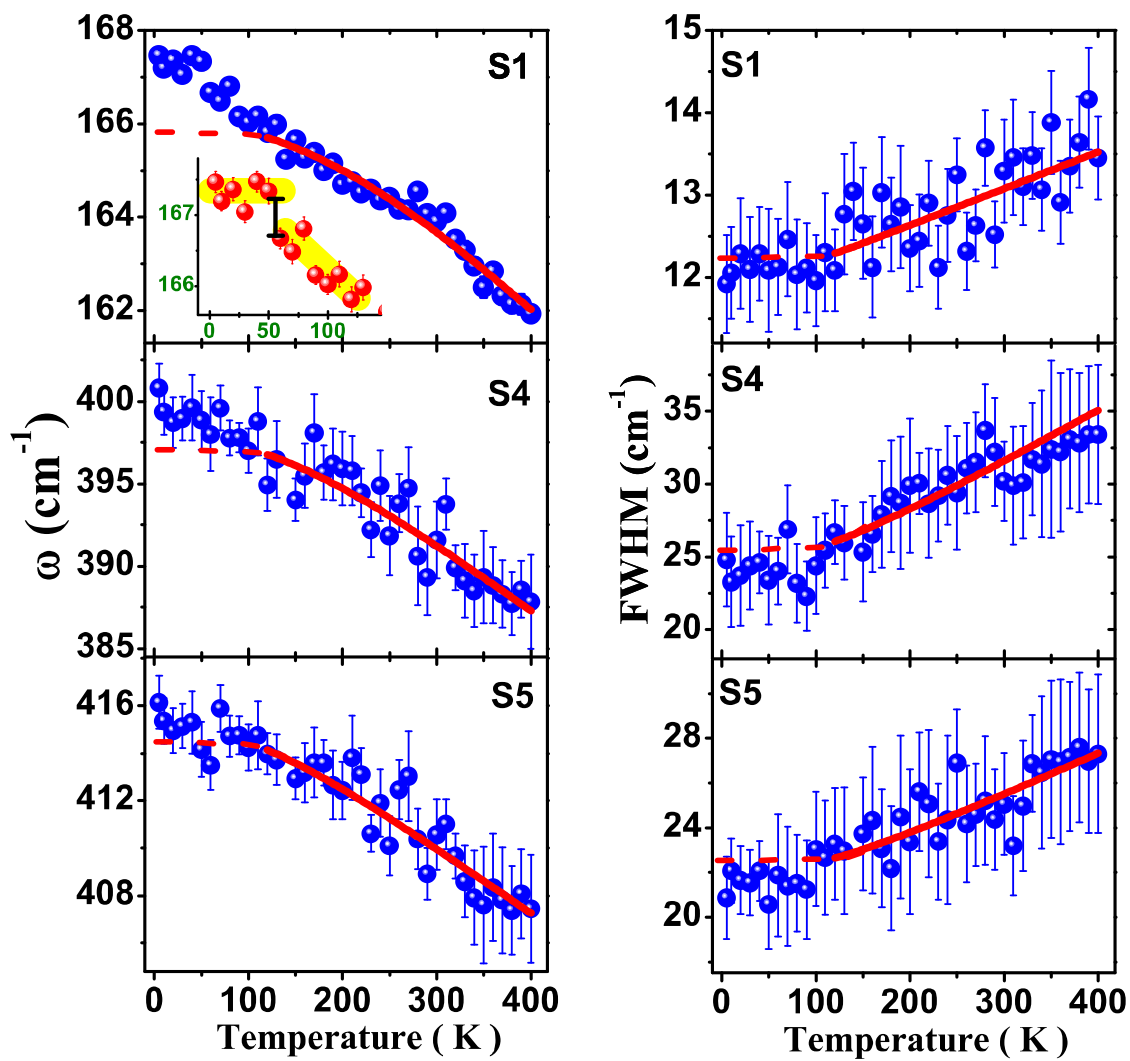

**Figure: 6**

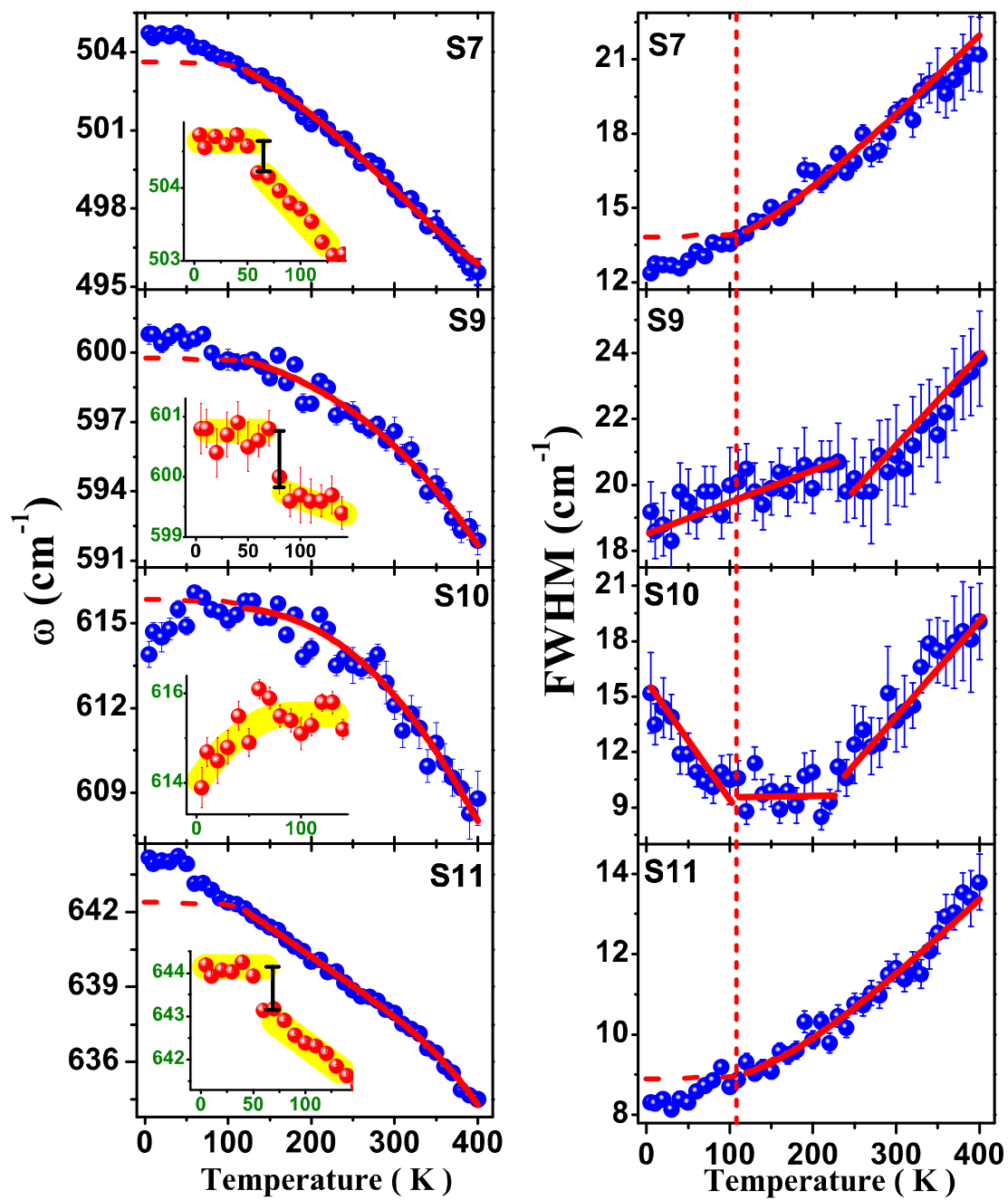

Figure: 7

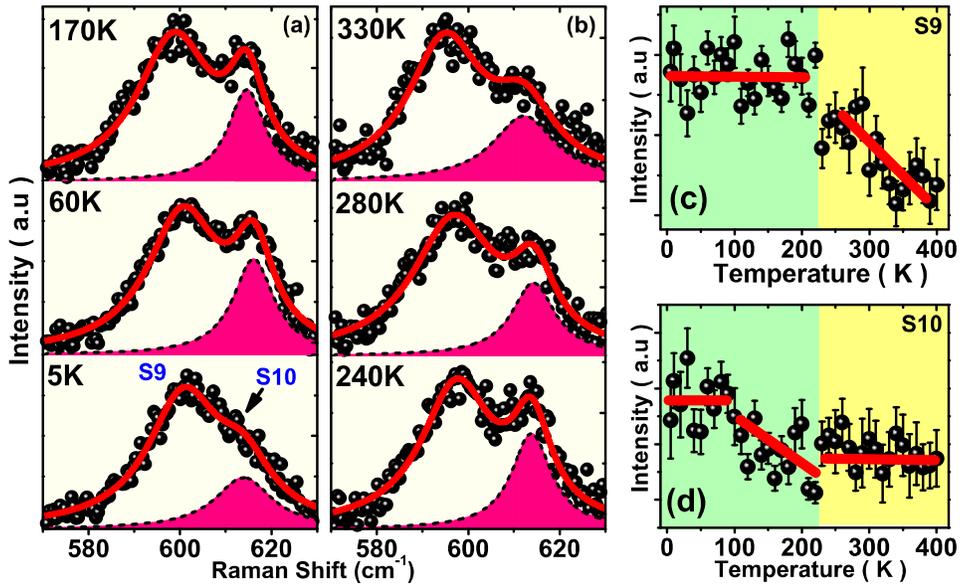

Figure: 8

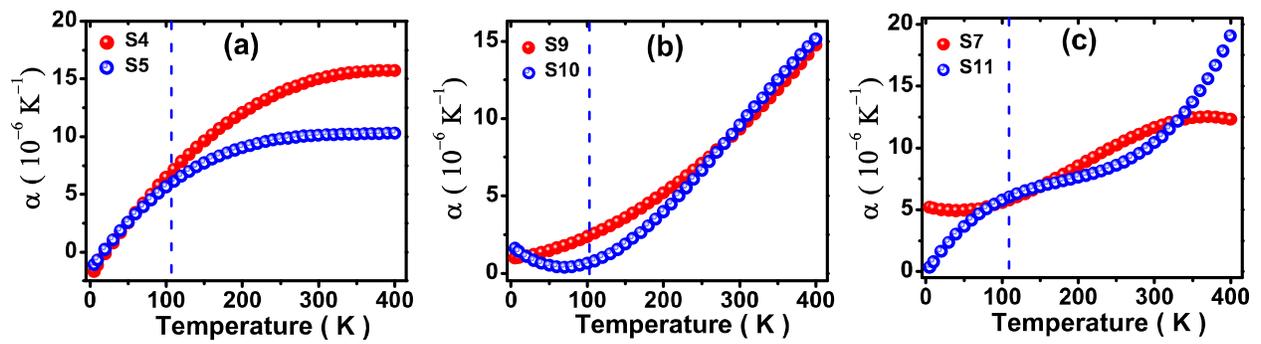

26